# Augmenting Supervised Emotion Recognition with Rule-Based Decision Model.

Amol S. Patwardhan and Gerald M. Knapp, *LSU*

*Abstract*—The aim of this research is development of rule based decision model for emotion recognition. This research also proposes using the rules for augmenting inter-corporal recognition accuracy in multimodal systems that use supervised learning techniques. The classifiers for such learning based recognition systems are susceptible to over fitting and only perform well on intra-corporal data. To overcome the limitation this research proposes using rule based model as an additional modality. The rules were developed using raw feature data from visual channel, based on human annotator agreement and existing studies that have attributed movement and postures to emotions. The outcome of the rule evaluations was combined during the decision phase of emotion recognition system. The results indicate rule based emotion recognition augment recognition accuracy of learning based systems and also provide better recognition rate across inter corpus emotion test data.

*Index Terms*— affect recognition, emotion recognition multimodal, rule-based recognition.

## I. INTRODUCTION

EMOTIONAL awareness in automated systems, computers and robotics greatly improves quality of interaction with humans [1]. For these interactions to be successful it is important that reliable emotion recognition systems exist. Emotions can be captured using audio-visual channels of input also known as modalities. In the past decade studies on multimodal emotion recognition have shown better accuracy compared to unimodal or bimodal emotion recognition [2], [3]. Researchers [4], [5], [6], [7], [8], [9], [10], [11], [12], [13] showed integrating various modalities for affect recognition not only provided better accuracy over individual modalities but also identified that hand gestures and body posture aided in emotion recognition. The multimodal systems use data from audio, visual and physiological channel to recognize emotions.

As a result, our implementation uses head, face, hand, body and speech for multimodal emotion recognition system instead of a unimodal system. Firstly, this study proposes development of rule based decision model to recognize emotions. Secondly the study proposes using the rules to augment learning based multimodal emotion recognition systems.

A. S. Patwardhan is a PhD student with Mechanical and Industrial Engineering Department of Louisiana State University, Baton Rouge, LA 70803 USA (e-mail: apatwa3@lsu.edu).

Dr. G. M. Knapp is with Mechanical and Industrial Engineering Department of Louisiana State University, Baton Rouge, LA 70803 USA (e-mail: gknapp@lsu.edu).

Many studies have successfully employed supervised learning techniques for emotion recognition. Results in [5] and classification techniques discussed in multimodal research surveys [2], [3] indicate performance improvement in affect recognition using support vector machine (SVM) classifiers. One of the limitations of these supervised learning based techniques is that they show high accuracy for only inter corpus data. These supervised learning techniques are also susceptible to over fitting. To overcome the limitations of using supervised learning alone, this research proposed using rule-based decision models to augment the learning based system accuracy. Research [14] has shown successful use of rule-based framework for arousal rating. This study focused on vocal features and proposed using rules as an alternative to supervised learning. Instead our research proposes using rule-based emotion estimation to augment the emotion recognition system.

Researchers [4], [15], [16], [17] have successfully used individual classification methods such as Bayesian classifiers, Hidden Markov Models (HMM) and SVM for affect recognition. On the other hand studies [18], [19], [20] have shown successful use of ensemble of classifiers, in speech based emotion recognition and face recognition. Hence we propose using the combination of SVM and rule based emotion models to form a hybrid multimodal emotion recognition system. This research argues that a combination of classifier and rule based recognition would improve precision and recall especially in case of multimodal affect recognition on data across corpuses other than those used for training the classifiers.

The motivation to use rules for emotion estimation was drawn from research done in emotion gesture recognition [21] and adaptive rule based facial expression recognitions [22]. The studies have demonstrated successful affect recognition by using limited set of gesture based rules and rules extracted from various facial expression profiles. Coulson [23] used computer generated mannequins from shoulder, hand, head descriptor and showed that each posture and movement can be attributed to one of the six basic emotions [24]. This study demonstrated that knowledge based rules for emotion recognition can be developed using annotator agreement.

Thus the contribution of our research is development of rules based on temporal (actor movement) and 3D (co-ordinate and skeleton joint) data in addition to purely static position based rules (rules that use body form) and using these rules to augment emotion recognition process. The temporal 3D data was captured from head movement, facial expressions, hand gestures and body posture. We also propose using the rules along with learning based system to improve inter-corporal recognition accuracy and generalizability of the multimodal emotion recognition system. In a survey done on multimodal systems [2], [3] the lack of clarity on whether some of the implementations are generalizable and tested against multiple datasets is discussed. We concur with this shortcoming and intend to evaluate our implementation against 3D data sets such as Microsoft Research Cambridge 12 (MSRC-12) [25], UCFKinect [26] and MSR Action 3D [27] dataset.

## II. OVERVIEW

In this research, we implemented a multimodal emotion recognition system using infrared sensor from Microsoft called Kinect. The 3D data from face, head, hand gestures and body movement was used for the visual channel. We used the openEar toolkit [28] for capturing audio data. The data from the various modalities was combined at the decision level. The classifiers for each modality were trained using SVM supervised learning technique. This research used a data mining tool called Weka [29] for training the model.

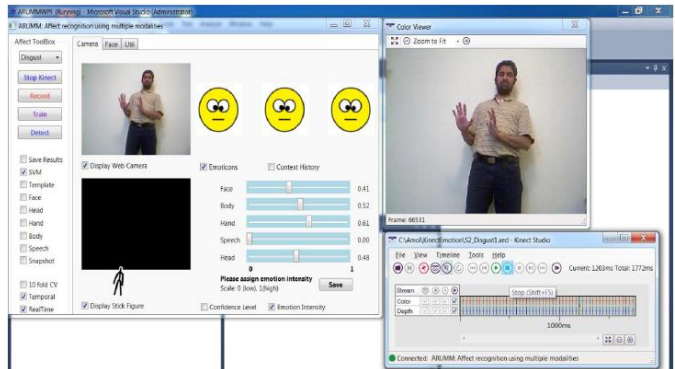
Fig. 2. Implementation of multimodal affect recognition system.

The objectives of this research are: 1) To show that knowledge based rules can be developed using human annotator agreement from temporal and 3D data from the visual channel and 2) To show that the emotion recognition accuracy of system can be augmented by using the rule based decision model in combination with supervised learning technique. Before we could evaluate it was necessary to measure the baseline accuracy numbers by using only the SVM supervised learning technique. The six basic emotions which are anger, surprise, disgust, sad, happy and fear were used as the candidate emotions for recognition process.

Each of the six emotions was enacted by 15 different individuals. The age of the participants was between 25 and 45 years. 5 participants were female and 10 participants were male. 5 participants were Americans and 10 participants were Asians. All the experiments were conducted in controlled lighting conditions and fully frontal position. The distance between the sensor and the participant was 1.5 to 4 meters.

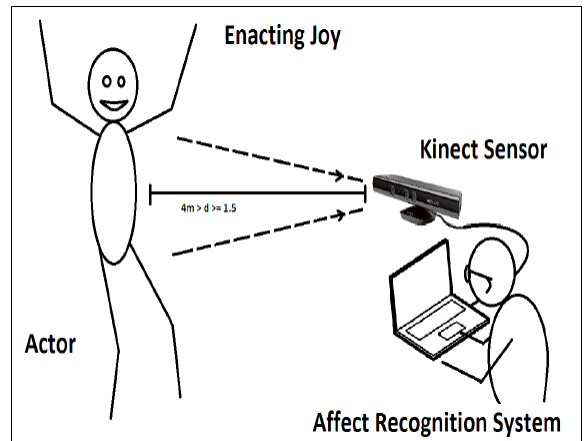
Fig. 1. A study participant enacting happiness in front of Kinect sensor.

Researchers [21], [22], [23] have shown that static head, hand and body positions can be used to develop rules using human annotator agreement. In addition to rules based on position of hands and body posture we developed temporal rules based on movement of head, facial expressions, hand and body. Once these rules were created we implemented the rules in the multimodal emotion recognition system and used the outcome of the rule decision model as an additional vote in the decision level fusion. The rule-based emotion outcome was used during decision level fusion along with results from classifier to predict the final emotion. The emotion recognition accuracy obtained using combination of rule based decision model and SVM was then compared with the baseline numbers obtained using only the supervised learning technique.

## III. FEATURE SELECTION

Facial features were extracted by the face recognition application programming interface (API) available in the Kinect software development kit (SDK). We used 60 non rigid features out of the 121 features. The intuition behind the initial selection of features was that only the features from the expressive part of the face were considered.

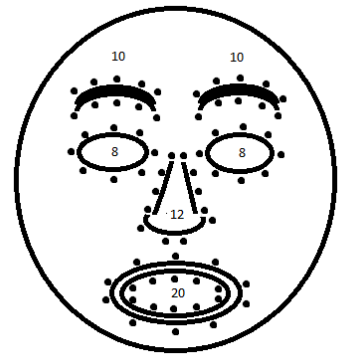
Fig. 3. Tracked features using face tracking API. The 60 features were chosen from the expressive part of the face.

The 60 non rigid features included x, y, z co-ordinates of the eyes, nose, lips, eye-lids, chin, cheek, forehead. In addition to the co-ordinates we also calculated the distance between each pair of feature and the angle made by each pair with the horizontal axis. The movement of each of this feature was captured for a window of 5 seconds which resulted in 100 frames. The velocity and displacement of each feature was calculated and used as temporal features. The features were stored in a format called arff used by the Weka data mining tool.

For tracking the head position and movement, 12 features along the border of the skull out of the 121 extracted features were chosen. The intuition behind these features was that the features were chosen such that they which would define the shape of the head as well as capture the movement such as pitch, yaw, roll, nod, shake, lateral, backward and forward motion of the head. Additionally, the distance between each pair of the feature, angle with the horizontal and movement of features across 100 frames was calculated.

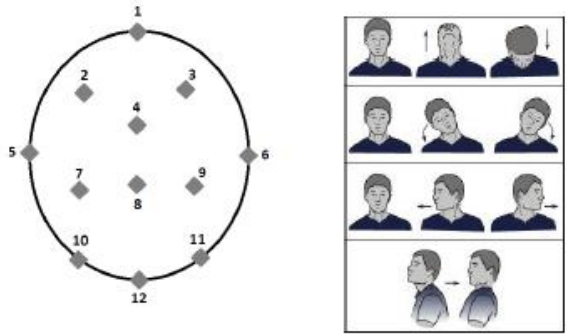

Fig. 4. Tracked features for head position and movement.

In case of hand gestures, palms, wrist, elbow and shoulder joints of both hands were tracked resulting in 8 features. These features were selected because using these features could be used to capture the vigorous movement of arms along all three axes. The distance between each pair of the joint, angle with the horizontal and velocity and displacement of each joint across 100 frames was calculated to create a feature vector.

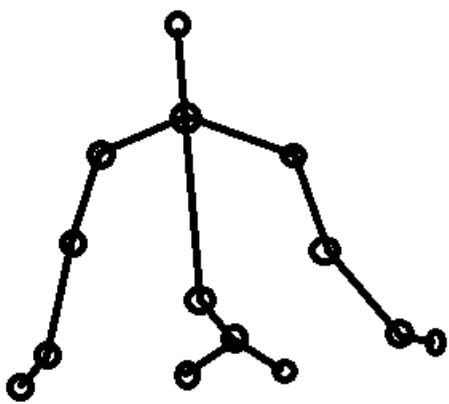

Fig. 5. Tracked features for hand and body position and movement.

For the body posture the center of spine, hip, left and right hip joints were tracked in addition to the joints of hand. The feature vector was constructed using distance between pair of joints, angle with horizontal and velocity and displacement of each joint across 100 frames. For the audio modality the openEar toolkit was used to extract the features and the pre-built SVM based classifiers were used for emotion recognition.

## IV. METHODOLOGY

15 individuals enacted six basic emotions in front of the Kinect sensor. The subjects were given a list of actions and dialogs to perform. This list of actions was prepared based on ideas from existing research [2], [3] and annotator agreement. 15 annotators were asked to label each action from the list to one of the 6 basic emotions. The author explained the situation to the actors and the actors reacted with one of the actions and dialog from the available list. The actors were given freedom to spontaneously enact actions and dialogs from the list or even improvise. The list of actions and the corresponding emotion based on manual annotator agreement is as follows:

TABLE 1
LIST OF POSES AND ACTIONS FOR ENACTED EMOTIONS

| Emotion | Action + Annotator agreement percentage | |
|---|---|---|
| Anger | Throwing an object. | 100% |
| | Punching. | 100% |
| | Holding head in frustration. | 80% |
| | Folding hands. | 70% |
| | Holding hands on waist. | 70% |
| | Moving forward threateningly. | 90% |
| | Throwing a fit. | 90% |
| | Raising arms in rage. | 70% |
| | Moving around in aggressively. | 70% |
| | Shouting in rage. | 100% |
| | Pointing a finger at someone. | 70% |
| | Threaten someone. | 90% |
| | Scowl. | 100% |
| Happy | Jumping in joy. | 100% |
| | Fist pumping in joy. | 100% |
| | Raising arms in air in happiness. | 100% |
| | Laughing. | 100% |
| | Smiling. | 100% |
| Disgust | Moving side-ways with arms evading. | 80% |
| | Holding nose. | 100% |
| | Looking down expressing disgust. | 100% |
| | Moving back in disgust. | 70% |
| Sad | Looking down leaning against wall. | 80% |
| | Looking down with hands on waist. | 70% |
| | Looking down with hands folded. | 80% |
| | Crying. | 100% |
| | Tooth ache. | 90% |
| | Head hurt. | 80% |
| Surprise | Raising arms in surprise. | 80% |
| | Moving back in surprise. | 90% |
| | Walking forward and getting startled. | 100% |
| | Holding arms near chest in surprise. | 90% |
| | Covering mouth with hands. | 90% |
| Fear | Moving backwards trying to evade. | 100% |
| | Moving sideways. | 80% |
| | Looking up and run away. | 90% |
| | Getting rid of an insect on shirt. | 90% |
| Neutral | Stand straight with no facial expression. | 100% |

The classifier for each modality was trained using SVM

learning technique. The data was split into 80% training and 20% test data. The model for the training data obtained from enacted emotions was built using 10-fold cross validation and Radial basis function as the non-linear kernel function. The parameters used for SVM training were as follows:

TABLE 2
SVM PARAMETERS FOR BASELINE

| Modality | C | Gamma |
|----------|---|-------|
| Head | 1 | 1/12 |
| Face | 1 | 1/60 |
| Hand | 1 | 1/8 |
| Body | 1 | 1/16 |

Research [5], [17] has shown that majority voting provides high accuracy results for multimodal emotion recognition. Hence for our multimodal emotion recognition implementation, a majority voting strategy was used. The results of classification from each modality were stored in a 2 dimensional buffer of size 10 x 6, where 10 is the number of result buffer instances and 6 is the number of modalities. The 6th modality was vote from rule based decision model. The total votes for each emotion were calculated and the emotions were ranked. This was done for 10 consecutive instances of result buffer and the emotion with the most votes was chosen as the final emotion detected by the system.

Based on experimental results 10 consecutive instances of results buffer provided the best multimodal results when the buffer size of changed between 5 and 25 in increments of 5. This event based and buffer dependent voting scheme was useful for accounting missing data, difference in time scales and missing vote from certain modalities at a given instance of time. The table below demonstrates the calculation of final emotion based on votes from different modalities.

TABLE 3
MULTIMODAL EMOTION RECOGNITION AT DECISION LEVEL

|  | T1 | T2 | T3 | …. | T9 | T10 | Final Prediction |
|--|----|----|----|-----|----|-----|------------------|
| Face | 1 | 1 | 3 |  | - | 4 |  |
| Head | 1 | 1 | 3 |  | - | 4 |  |
| Body | 4 | 4 | 4 |  | 3 | 3 |  |
| Hand | 4 | 4 | 3 |  | 4 | 3 |  |
| Speech | - | - | - |  | 4 | 4 |  |
| Rule | 4 | 4 | 3 |  | 4 | - |  |
| Emotion | 4 | 4 | 3 |  | 4 | 4 | 4 |

The columns represent instances of result buffer when a vote was available from various modalities. The class labels used for the 6 basic emotions were Anger = 0, Happiness = 1, Surprise = 2, Disgust = 3, Fear = 4, Sad = 5, Neutral = 6 and Unavailable = -. Thus based on votes from T1 column, Fear received most votes. Similarly based on votes from each column Fear received the most votes after 10 counts of vote. Thus the system predicted final emotion was Fear.

Further analysis of the results buffer shows that the data for head and face was available from the same depth frame (data containing 3d co-ordinates) of the Kinect sensor. As a result, the result of emotion recognition was unavailable for both head and face at the same time and more often. This was because of the course features on the face requiring more processing time and the lighting conditions, movement and orientation of the face caused generation of fewer reliable frames from the face input.

The same correlation could be observed in case of features and emotion results from the hand and body channel. Similarly input from sound modality was not always available because the actor would not utter sound during the entire duration of the emotional act. For the baseline experiment (one with only the SVM based implementation) the vote from rule based decision model was switched off and thus excluded.

The SVM only baseline multimodal emotion recognition results are as shown below:

TABLE 4
MULTIMODAL EMOTION RECOGNITION USING SVM

| 0 | 1 | 2 | 3 | 4 | 5 | 6 |  |  |
|---|---|---|---|---|---|---|--|--|
| 378 | 3 | 0 | 33 | 0 | 49 | 0 | 0 | Anger |
| 47 | 426 | 2 | 18 | 17 | 0 | 0 | 1 | Happy |
| 15 | 16 | 447 | 0 | 30 | 0 | 0 | 2 | Surprise |
| 92 | 25 | 1 | 328 | 17 | 9 | 0 | 3 | Disgust |
| 16 | 25 | 15 | 12 | 481 | 0 | 0 | 4 | Fear |
| 33 | 10 | 0 | 13 | 0 | 459 | 1 | 5 | Sad |
| 0 | 0 | 0 | 0 | 2 | 2 | 465 | 6 | Neutral |

## V. RULE BASED DECISION MODEL

The research aim is to develop emotion recognition rules from raw 3d static and temporal data which would serve as a decision model. To develop these rules 15 participants were asked to annotate static images and video clips to an emotion class.

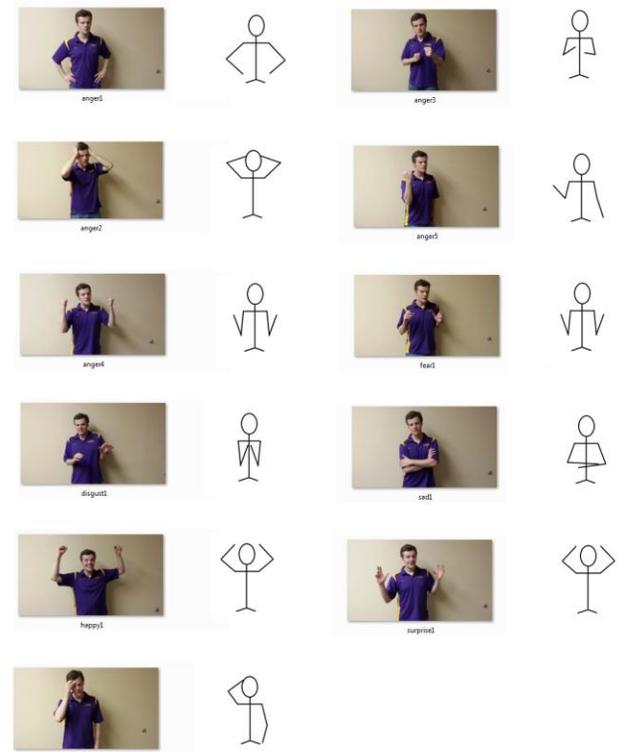

Fig. 6. Tracked features for head position and movement.

The static images contained posed emotions and were used for creating rules based on position and body form. The video clips containing actions were used to create rules based on body movement. Each image mapped to a set of rules defined using co-ordinates, angle and distance of the features from various modalities. Similarly, each video clip was mapped to a set of rules defined using movement along certain axis, frequency of movement, velocity and displacement of the features discussed earlier.

Thus the rules captured not only static position but also temporal gestures, postures and movements of body parts used in expressing emotions. The use of images and video clips simplified the annotation process because the annotators did not have to analyze each rule associated with the emotion. The table below shows an illustrative list of rules developed for emotion recognition.

TABLE 5
RULE DESCRIPTORS FOR EMOTION ESTIMATION

| Rule ID | Rule Descriptor |
|---------|-----------------|
| R1 | Angle of left elbow |
| R2 | Angle of right elbow |
| R3 | Angle between left shoulder and arm |
| R4 | Angle between right shoulder and arm |
| R5 | Angle of spine |
| R6 | Angle of head |
| R7 | Y co-ordinate of wrist > Y co-ordinate of elbow |
| R8 | Y co-ordinate of elbow > Y co-ordinate of shoulder |
| R9 | X co-ordinate of wrist > X co-ordinate of elbow |
| R10 | X co-ordinate of elbow > X co-ordinate of shoulder |
| R11 | Z co-ordinate of wrist > Z co-ordinate of elbow |
| R12 | Z co-ordinate of elbow > Z co-ordinate of shoulder |
| R13 | X co-ordinate of wrist < X co-ordinate of shoulder |
| R14 | Frequency of head nod |
| R15 | Frequency of waving hand |
| R16 | Frequency of forward movement |
| R17 | Frequency of backward movement |
| R18 | Frequency of sideways movement |
| R19 | Frequency of shaking head sideways |
| R20 | Distance between eyebrow and eyes |
| R21 | Distance between upper and lower lip |
| R22 | Distance between nose tip and upper lip |
| R23 | Distance between corners of lip |
| R24 | Distance between upper and lower eyelid |

Once the annotation was complete the minimum and maximum value of each angle, distance and velocity was calculated to create a threshold for each rule. For instance, the minimum angle at the elbow for an angry pose with both hands on the waist was 92 degrees while the maximum was 95 degrees.

Based on human annotator agreement results and the threshold values, the rules were implemented using the raw 3D features and fed to the multimodal system as if it were a separate modality. Thus the outcome of the rule evaluation counted as a vote during the decision level fusion and emotion recognition process. The results of multimodal emotion recognition using combination of learning based technique and rules are shown below.

TABLE 6
MULTIMODAL EMOTION RECOGNITION USING SVM + RULES

| 0 | 1 | 2 | 3 | 4 | 5 | 6 | | |
|---|---|---|---|---|---|---|---|---|
| 412 | 5 | 15 | 12 | 0 | 19 | 0 | 0 | Anger |
| 34 | 431 | 28 | 9 | 8 | 0 | 0 | 1 | Happy |
| 13 | 17 | 459 | 3 | 16 | 0 | 0 | 2 | Surprise |
| 78 | 21 | 6 | 339 | 17 | 11 | 0 | 3 | Disgust |
| 7 | 17 | 23 | 16 | 486 | 0 | 0 | 4 | Fear |
| 24 | 7 | 1 | 10 | 2 | 471 | 1 | 5 | Sad |
| 0 | 0 | 0 | 0 | 0 | 12 | 457 | 6 | Neutral |

The above results indicate an improvement in the recognition accuracy when rules were used in combination with learning based technique compared to the accuracy obtained using SVM only. In order to test the effectiveness of the approach on inter-corpus test data we used MSRC-12 dataset, UCFKinect dataset and MSR Action 3D dataset.

These datasets are not directly annotated with one of the 6 basic emotions. The format of the features is also different from our feature definitions. To overcome this limitation to test our recognition system against these datasets we first mapped each activity in the dataset to an emotion class.

We asked 15 participants to label each activity to an emotion class. The results of the mapping between each activity from the dataset and an emotion class are shown in the table below:

TABLE 7
MAPPING FOR MSRC-12

| Action | Emotion/Annotator agreement |
|--------|------------------------------|
| Crouch or hide | Fear 100% |
| Shoot with a pistol | Angry 90% |
| Throw an object | Angry 90% |
| Change weapon | Inconclusive |
| Kick to attack an enemy | Angry 100% |
| Put on night vision goggle | Inconclusive |
| Protest the music | Angry 80% |
| Music based gestures | Inconclusive |

TABLE 8
MAPPING FOR UCFKINECT

| Action | Emotion/Annotator agreement |
|--------|------------------------------|
| Balance, climb ladder, climb up | Inconclusive |
| Duck | Fear 100% |
| Hop | Surprise 60% |
| Kick | Anger 80% |
| Leap | Surprise 70% |
| Punch | Anger 100% |
| Run | Fear 80% |
| Step back | Fear 80%, Disgust 60% |
| Step front | Anger 60% |
| Step left | Disgust 70% |
| Step right | Disgust 70% |
| Turn left, Turn right, Vault | Inconclusive |

TABLE 9
MAPPING FOR MSRACTION

| Action | Emotion/Annotator agreement |
|--------|----------------------------|
| Sit down, stand up, | Inconclusive |
| Hand clapping | Happy 100% |
| Hand waving | Happy 70% |
| Cheer up | Happy 100% |
| Boxing | Anger 100% |
| Toss a paper | Anger 90% |

Based on the results of the mapping we created a version of each dataset so that it could be tested with our emotion recognition implementation. The mapping indicates that not all emotions were adequately represented in the datasets. This is justified because the MSRC-12, UCFKinect and MSRAction 3D datasets are intended for human activities and not specifically for emotion representation.

5 participants enacted the list of actions which were annotated with the emotion class using the action to emotion mapping. The 3D temporal data was then supplied to the multimodal emotion recognition system and experiments were conducted using only SVM and then using combination of SVM and rule based decision model. The results of the SVM based emotion recognition are shown below.

TABLE 10
SVM ON MSRC-12 DATASET

| 0 | 1 | 2 | 3 | 4 | 5 | 6 | | |
|---|---|---|---|---|---|---|---|---|
| 2339 | 281 | 196 | 31 | 27 | 10 | 0 | 0 | Anger |
| - | - | - | - | - | - | - | 1 | Happy |
| - | - | - | - | - | - | - | 2 | Surprise |
| - | - | - | - | - | - | - | 3 | Disgust |
| 15 | 22 | 178 | 23 | 2831 | 48 | 0 | 4 | Fear |
| - | - | - | - | - | - | - | 5 | Sad |
| - | - | - | - | - | - | - | 6 | Neutral |

TABLE 11
SVM ON UCFKINECT DATASET

| 0 | 1 | 2 | 3 | 4 | 5 | 6 | | |
|---|---|---|---|---|---|---|---|---|
| 321 | 86 | 53 | 0 | 20 | 0 | 0 | 0 | Anger |
| - | - | - | - | - | - | - | 1 | Happy |
| 73 | 56 | 385 | 4 | 0 | 2 | 0 | 2 | Surprise |
| 4 | 0 | 23 | 365 | 44 | 12 | 0 | 3 | Disgust |
| 0 | 0 | 17 | 38 | 359 | 5 | 3 | 4 | Fear |
| - | - | - | - | - | - | - | 5 | Sad |
| - | - | - | - | - | - | - | 6 | Neutral |

TABLE 12
SVM ON MSRACTION DATASET

| 0 | 1 | 2 | 3 | 4 | 5 | 6 | | |
|---|---|---|---|---|---|---|---|---|
| 258 | 68 | 13 | 10 | 0 | 5 | 0 | 0 | Anger |
| 59 | 291 | 25 | 4 | 2 | 0 | 0 | 1 | Happy |
| - | - | - | - | - | - | - | 2 | Surprise |
| - | - | - | - | - | - | - | 3 | Disgust |
| - | - | - | - | - | - | - | 4 | Fear |
| - | - | - | - | - | - | - | 5 | Sad |
| - | - | - | - | - | - | - | 6 | Neutral |

After the experiments using SVM based emotion recognition, the same set of data was tested using a combination of SVM and rule based decision model. The results of the experiments are shown below.

TABLE 13
SVM + RULE ON MSRC-12 DATASET

| 0 | 1 | 2 | 3 | 4 | 5 | 6 | | |
|---|---|---|---|---|---|---|---|---|
| 2573 | 176 | 107 | 20 | 8 | 0 | 0 | 0 | Anger |
| - | - | - | - | - | - | - | 1 | Happy |
| - | - | - | - | - | - | - | 2 | Surprise |
| - | - | - | - | - | - | - | 3 | Disgust |
| 8 | 25 | 140 | 38 | 2889 | 17 | 0 | 4 | Fear |
| - | - | - | - | - | - | - | 5 | Sad |
| - | - | - | - | - | - | - | 6 | Neutral |

TABLE 14
SVM + RULE ON UCFKINECT DATASET

| 0 | 1 | 2 | 3 | 4 | 5 | 6 | | |
|---|---|---|---|---|---|---|---|---|
| 362 | 65 | 29 | 15 | 7 | 2 | 0 | 0 | Anger |
| - | - | - | - | - | - | - | 1 | Happy |
| 59 | 34 | 412 | 11 | 4 | 0 | 0 | 2 | Surprise |
| 0 | 0 | 27 | 389 | 15 | 17 | 0 | 3 | Disgust |
| 7 | 0 | 11 | 8 | 392 | 4 | 0 | 4 | Fear |
| - | - | - | - | - | - | - | 5 | Sad |
| - | - | - | - | - | - | - | 6 | Neutral |

TABLE 15
SVM + RULE ON MSRACTION DATASET

| 0 | 1 | 2 | 3 | 4 | 5 | 6 | | |
|---|---|---|---|---|---|---|---|---|
| 283 | 45 | 21 | 5 | 0 | 0 | 0 | 0 | Anger |
| 25 | 326 | 18 | 12 | 0 | 0 | 0 | 1 | Happy |
| - | - | - | - | - | - | - | 2 | Surprise |
| - | - | - | - | - | - | - | 3 | Disgust |
| - | - | - | - | - | - | - | 4 | Fear |
| - | - | - | - | - | - | - | 5 | Sad |
| - | - | - | - | - | - | - | 6 | Neutral |

The tables 16 to 19 show the precision, recall, F-score and accuracy calculated for only learning based emotion recognition. Tables 20 to 23 show the same calculations for emotion recognition using the combination of learning based and rules.

TABLE 16
SVM (BASELINE)

| Precision | Recall | F score | Accuracy | | |
|-----------|--------|---------|----------|---|---|
| 0.651 | 0.817 | 0.725 | 0.167 | 0 | Anger |
| 0.844 | 0.836 | 0.84 | 0.145 | 1 | Happy |
| 0.962 | 0.88 | 0.919 | 0.134 | 2 | Surprise |
| 0.812 | 0.695 | 0.749 | 0.116 | 3 | Disgust |
| 0.88 | 0.877 | 0.878 | 0.157 | 4 | Fear |
| 0.885 | 0.89 | 0.887 | 0.149 | 5 | Sad |
| 0.998 | 0.992 | 0.995 | 0.134 | 6 | Neutral |

TABLE 17
SVM ON MSRC-12 DATASET

| Precision | Recall | F score | Accuracy | | |
|-----------|--------|---------|----------|---|---|
| 0.994 | 0.812 | 0.894 | 0.393 | 0 | Anger |
| - | - | - | - | 1 | Happy |

| Precision | Recall | F score | Accuracy | | |
|-----------|--------|---------|----------|---|---------|
| - | - | - | - | 2 | Surprise |
| - | - | - | - | 3 | Disgust |
| 0.991 | 0.909 | 0.948 | 0.477 | 4 | Fear |
| - | - | - | - | 5 | Sad |
| - | - | - | - | 6 | Neutral |

TABLE 18
SVM ON UCFKINECT DATASET

| Precision | Recall | F score | Accuracy | | |
|-----------|--------|---------|----------|---|---------|
| 0.807 | 0.669 | 0.732 | 0.213 | 0 | Anger |
| - | - | - | - | 1 | Happy |
| 0.806 | 0.741 | 0.772 | 0.256 | 2 | Surprise |
| 0.897 | 0.815 | 0.854 | 0.218 | 3 | Disgust |
| 0.849 | 0.851 | 0.85 | 0.227 | 4 | Fear |
| - | - | - | - | 5 | Sad |
| - | - | - | - | 6 | Neutral |

TABLE 19
SVM ON MSRACTION DATASET

| Precision | Recall | F score | Accuracy | | |
|-----------|--------|---------|----------|---|---------|
| 0.814 | 0.729 | 0.77 | 0.432 | 0 | Anger |
| 0.811 | 0.764 | 0.787 | 0.489 | 1 | Happy |
| - | - | - | - | 2 | Surprise |
| - | - | - | - | 3 | Disgust |
| - | - | - | - | 4 | Fear |
| - | - | - | - | 5 | Sad |
| - | - | - | - | 6 | Neutral |

TABLE 20
SVM + RULE (BASELINE)

| Precision | Recall | F score | Accuracy | | |
|-----------|--------|---------|----------|---|---------|
| 0.726 | 0.89 | 0.8 | 0.163 | 0 | Anger |
| 0.866 | 0.846 | 0.856 | 0.143 | 1 | Happy |
| 0.863 | 0.904 | 0.883 | 0.153 | 2 | Surprise |
| 0.872 | 0.719 | 0.788 | 0.112 | 3 | Disgust |
| 0.919 | 0.886 | 0.902 | 0.152 | 4 | Fear |
| 0.919 | 0.913 | 0.916 | 0.148 | 5 | Sad |
| 0.998 | 0.975 | 0.986 | 0.132 | 6 | Neutral |

TABLE 21
SVM + RULE ON MSRC-12 DATASET

| Precision | Recall | F score | Accuracy | | |
|-----------|--------|---------|----------|---|---------|
| 0.997 | 0.893 | 0.942 | 0.431 | 0 | Anger |
| - | - | - | - | 1 | Happy |
| - | - | - | - | 2 | Surprise |
| - | - | - | - | 3 | Disgust |
| 0.998 | 0.927 | 0.961 | 0.483 | 4 | Fear |
| - | - | - | - | 5 | Sad |
| - | - | - | - | 6 | Neutral |

TABLE 22
SVM + RULE ON UCFKINECT DATASET

| Precision | Recall | F score | Accuracy | | |
|-----------|--------|---------|----------|---|---------|
| 0.846 | 0.755 | 0.798 | 0.229 | 0 | Anger |
| - | - | - | - | 1 | Happy |
| 0.861 | 0.793 | 0.825 | 0.257 | 2 | Surprise |
| 0.92 | 0.869 | 0.894 | 0.227 | 3 | Disgust |
| 0.938 | 0.929 | 0.934 | 0.224 | 4 | Fear |
| - | - | - | - | 5 | Sad |
| - | - | - | - | 6 | Neutral |

TABLE 23
SVM + RULE ON MSRACTION DATASET

| Precision | Recall | F score | Accuracy | | |
|-----------|--------|---------|----------|---|---------|
| 0.919 | 0.8 | 0.855 | 0.42 | 0 | Anger |
| 0.879 | 0.856 | 0.868 | 0.505 | 1 | Happy |
| - | - | - | - | 2 | Surprise |
| - | - | - | - | 3 | Disgust |
| - | - | - | - | 4 | Fear |
| - | - | - | - | 5 | Sad |
| - | - | - | - | 6 | Neutral |

A comparison between the precision, recall of only learning based recognition and combination of rule based recognition shows that the precision and recall improves when rules are used to augment the multimodal emotion recognition system.

## VI. CONCLUSION

This research developed rules using 1) Static data that provided measurement of body form and feature co-ordinates, 2) Temporal data that provided information on actor facial expressions, hand gestures and body movement and 3) 3D data from infrared sensor depth and skeleton frames. The research captured the variety of popular emotional actions from the state of the art studies and coded them into emotion recognition rules. The rules were developed using human annotator agreement.

The results indicate that the rules can be used in multimodal emotion recognition systems and are useful in improving the accuracy of learning based system especially against inter-corporal data. As a future scope we intend to extend the set of rules and make the rule set available as a comprehensive reference for emotion recognition studies and also test it on newer Kinect emotion corpus as they become available.

## REFERENCES

[1] R. W. Picard and R. Picard, "*Affective Computing*", Cambridge, MIT Press. 1997.

[2] M. Pantic and L. J. Rothkrantz, "Toward an Affect-Sensitive Multimodal Human–Computer Interaction," *Proceedings of the IEEE, vol.91, no.1, pp. 1370-1390.2003.*

[3] Z. Zeng, M. Pantic, G.I. Roisman, and T.S. Huang, "A Survey of Affect Recognition Methods: Audio, Visual, and Spontaneous Expressions," IEEE Trans. Pattern Analysis and Machine Intelligence, vol. 31, no. 1, pp. 39-58, Jan. 2009

[4] S. C. Tan and A. Nareyek, "Integrating Facial, Gesture, and Posture Emotion Expression for a 3D Virtual Agent." *Proceedings of the* 14th International Conference on Computer Games: AI, Animation, Mobile, Interactive Multimedia, Educational & Serious Games, pp. 23-31.2009.

[5] G. Castellano, L. Kessous, and G. Caridakis, Multimodal Emotion Recognition from Expressive Faces, Body Gestures and Speech," *Proceedings of* 2nd International Conference on Affective Computing and Intelligent Interaction, vol. 247, pp 375-388. 2007.

[6] S. Emerich, E. Lupu, and A. Apatean, "Bimodal approach in emotion recognition using speech and facial expressions." *International Symposium on* Signals, Circuits and Systems, pp. 1-4. 2009.

[7] L. Chen, T. Huang, T. Miyasato, and R. Nakatsu, "Multimodal Human Emotion/Expression Recognition," *Proc. Third IEEE Int'l Conf.* Automatic Face and Gesture Recognition, pp. 366-371, 1998.

[8] K. Scherer and H. Ellgring, "Multimodal Expression of Emotion:

Affect Programs or Componential Appraisal Patterns?," Emotion, vol. 7, pp. 158-171, 2007.

[9] A. Kapoor and R.W. Picard, "Multimodal Affect Recognition in Learning Environments," *Proc. 13th Ann. ACM Int'l Conf.* Multimedia, pp. 677-682, 2005.

[10] S. D'Mello and A. Graesser, "Multimodal Semi-Automated Affect Detection from Conversational Cues, Gross Body Language, and Facial Features," User Modeling and User-Adapted Interaction, vol. 10, pp. 147-187, 2010.

[11] T. Baenziger, D. Grandjean, and K.R. Scherer, "Emotion Recognition from Expressions in Face, Voice, and Body. The Multimodal Emotion Recognition Test (MERT)," Emotion, vol. 9, pp. 691-704, 2009.

[12] C. Busso et al., "Analysis of Emotion Recognition Using Facial Expressions, Speech and Multimodal Information," *Proc. Int'l Conf.* Multimodal Interfaces, T.D.R. Sharma, M.P. Harper, G. Lazzari, and M. Turk, eds., pp. 205-211, 2004.
N. Sebe, I. Cohen, and T.S. Huang, "Multimodal Emotion Recognition," Handbook of Pattern Recognition and Computer Vision, World Scientific, 2005.

[13] R. Cowie, E. Douglas-Cowie, N. Tsapatsoulis, G. Votsis, S. Kollias, W. Fellenz, and J. Taylor, "Emotion Recognition in Human-Computer Interaction," IEEE Signal Processing Magazine, vol. 18, no. 1, pp. 32-80, 2001.

[14] D. Bone, C. Lee, S. Narayan, "Robust Unsupervised Arousal Rating: A Rule-Based Framework with Knowledge-Inspired Vocal Features." IEEE Transactions on Affective Computing, vol. 5, no. 2. 2014.

[15] Z. Zeng, T. Jilin, B. M. Pianfetti, T. S, Huang, "Audio–Visual Affective Expression Recognition through Multistream Fused HMM." IEEE Transactions on Multimedia, vol. 10, no. 4, pp. 570-577. 2008.

[16] K. Takahashi, "Remarks on SVM-based emotion recognition from multi-modal bio-potential signals," *13th IEEE International Workshop on Robot and Human Interactive Communication*, pp. 95-100, 2004.

[17] M. F. Valstar, H. Gunes, and M. Pantic, "How to Distinguish Posed from Spontaneous Smiles using Geometric Features." *In Proceedings of ACM International Conference on* Multimodal Interfaces, pp. 38-45. 2007.

[18] T. Danisman, A. Alpkocak, "Emotion Classification of Audio Signals Using Ensemble of Support Vector Machines." *Proceedings of 4th* IEEE Tutorial and Research Workshop on Perception and Interactive Technologies for Speech-Based Systems, pp. 205-216. 2008.

[19] Y. Liu, Y. Zheng, and Y. Chen, "Ensemble classification based on correlation analysis for face recognition." Neural Networks, vol. 299, no. 303, pp. 1-8. 2008.

[20] B. Schuller, S. Reiter, R. Müller, M. Al-Hames, M. Lang, and G. Rigoll, "Speaker Independent Speech Emotion Recognition by Ensemble Classification." *IEEE International Conference on* Multimedia and Expo, pp. 864-867. 2005.

[21] L. Zhang and B. Yap, "Affect Detection from Text-Based Virtual Improvisation and Emotional Gesture Recognition." Advances in Human-Computer Interaction, 2012.

[22] S. Ioannou, A. Raouzaiou, K. Karpouzis, M. Pertselakis, N. Tsapatsoulis, and S. Kollias, "Adaptive rule-based facial expression recognition." *Lecture Notes in* Artificial Intelligence, vol. 3025, pp. 466-475. 2004.

[23] M. Coulson, "Attributing emotion to static body postures: recognition accuracy, confusions, and viewpoint dependence." Journal of Nonverbal Behavior, vol. 28, no. 2, pp. 117-139. 2010.

[24] P. Ekman, "An argument for basic emotions." *Cognition & emotion* vol. 6, no. 3-4, pp. 169-200. 1992.

[25] S. Fothergill, H. M. Mentis, P. Kohli, and S. Nowozin, "Instructing people for training gestural interactive systems" CHI, ACM, pp. 1737-1746. 2012.

[26] S. Z. Masood, C. Ellis, M. F. Tappen, J. J. LaViola Jr, and R. Sukthankar, "Exploring the Trade-off Between Accuracy and Observational Latency in Action Recognition," International Journal of Computer Vision, vol.10, no. 3, 2010.

[27] J. Yuang, Z. Liu, Y. Wu, "Discriminative Subvolume Search for Efficient Action Detection," *IEEE Conference on* Computer Vision and Pattern Recognition (CVPR 2009), pp. 22-24, 2009.

[28] F. Eyben, M. Wöllmer, B. Schuller, "openEAR - Introducing the Munich Open-Source Emotion and Affect Recognition Toolkit," *In Proc. 4th International HUMAINE* Association Conference on Affective Computing and Intelligent Interaction, IEEE, Amsterdam, The Netherlands, 2009.

[29] M. Hall, E. Frank, G. Holmes, B. Pfahringer, P. Reutemann, I. H. Witten, "The WEKA Data Mining Software: An Update" SIGKDD Explorations, vol. 11, no. 1, 2009.

[30] A. S. Patwardhan, 2016. "Structured Unit Testable Templated Code for Efficient Code Review Process", PeerJ Computer Science (in review), 2016.

[31] A. S. Patwardhan, and R. S. Patwardhan, "XML Entity Architecture for Efficient Software Integration", International Journal for Research in Applied Science and Engineering Technology (IJRASET), vol. 4, no. 6, June 2016.

[32] A. S. Patwardhan and G. M. Knapp, "Affect Intensity Estimation Using Multiple Modalities," Florida Artificial Intelligence Research Society Conference, May. 2014.

[33] A. S. Patwardhan, R. S. Patwardhan, and S. S. Vartak, "Self-Contained Cross-Cutting Pipeline Software Architecture," International Research Journal of Engineering and Technology (IRJET), vol. 3, no. 5, May. 2016.

[34] A. S. Patwardhan, "An Architecture for Adaptive Real Time Communication with Embedded Devices," LSU, 2006.

[35] A. S. Patwardhan and G. M. Knapp, "Multimodal Affect Analysis for Product Feedback Assessment," IIE Annual Conference. Proceedings. Institute of Industrial Engineers-Publisher, 2013.

[36] A. S. Patwardhan and G. M. Knapp, "Aggressive Action and Anger Detection from Multiple Modalities using Kinect", submitted to ACM Transactions on Intelligent Systems and Technology (ACM TIST) (in review).

[37] A. S. Patwardhan and G. M. Knapp, "EmoFit: Affect Monitoring System for Sedentary Jobs," preprint, arXiv.org, 2016.

[38] A. S. Patwardhan, J. Kidd, T. Urena and A. Rajagopalan, "Embracing Agile methodology during DevOps Developer Internship Program", IEEE Software (in review), 2016.

[39] A. S. Patwardhan, "Analysis of Software Delivery Process Shortcomings and Architectural Pitfalls", PeerJ Computer Science (in review), 2016.

[40] A. S. Patwardhan, "Multimodal Affect Recognition using Kinect", ACM TIST (in review), 2016.